# Thermally-operating regimes for a three-layer liquid metal battery

Riyad Hassaine, Olivier Crumeyrolle[*] and Innocent Mutabazi

Normandie Université, UNIHAVRE, LOMC, UMR 6294 CNRS-ULHN

53, rue de Prony, 76058 Le Havre Cedex, France

**Abstract**

Thermally-operating regimes of a three-layer liquid metal battery are determined in terms of the aspect ratio. The heat exchange between the battery container and the environment is taken into account. Velocity fields and temperature distribution of the convection state inside the battery are determined from numerical simulations. For a chosen geometry, the identified operating regime result in characteristic charge/discharge time close to half a day.

## I. Introduction

The decarbonisation of energy production has surged the development of techniques capable of harnessing the renewable energy from intermittent sources (wind turbines, solar panels) [1]. Liquid metal batteries (LMBs) represent one of the promising solutions for electrochemical storage of the produced energy before it can be integrated into the global electrical grid [1-3]. LMBs have gained a great interest as they are supposed to be safe and represent affordable solution for electrochemical storage that can modify the operating way of the electric grids. In that context, to buffer the daily electricity demand is of great interest. Since more than a decade, many studies have been performed to investigate the chemical and physical processes in LMB [1-8]. Kelley and Weier [9] have made a review of most of the hydrodynamic phenomena that can be encountered in LMBs. Indeed, many phenomena are driven in LMB, the most important being the thermal convection generated by internal heating due to the applied current in the electrolyte [5], the electrothermal and electrochemical reactions [10], the magnetic convection [11] and the Tayler instability [12] due to the magnetic field and the electro-vortex flow due to the Lorentz force acting in the LMB by applied external magnetic and electric fields [13].

[*] Corresponding author: olivier.crumeyrolle@univ-lehavre.fr

Thermal convection induced by internal heating in a simple layer of liquid has been investigated by many authors [14-18]. Thorough investigations of thermal convection in LMB were recently performed by Shen and Zikanov [5] in a symmetrical LMB in which the properties of both the electrodes were supposed identical except for the thermal and electrical conductivities. Köllner *et al*. [7] have analyzed thermal convection and thermocapillary effects in an asymmetrical LMB. In both studies, the environment was supposed to impose its temperature to the container of the battery, i.e. isothermal Dirichlet boundary conditions were assumed. In the present work, we will illustrate our analyzis using the properties of the LMB investigated by Kim *et al*. [1] and Kelley & Weier [9]. The analyzis incorporates both the situation when the LMB container has the same temperature as the environment and the case when the heat transfer between the LMB container and the environment is hindered, as described by an exchange coefficient. As far as we know from available literature, the later situation has not been investigated in LMB. For a reasonable combination of air cooling, temperature and emissivity, an exchange coefficient of 10 W/(m$^2$·K) is achievable. Hence heat flux of 1 kW/m$^2$ will already result in temperature jump of 100 °C. Higher temperature jump would be obtained with radiation shielding / low emissivity surfaces. This helps to keep the LMB warm with low-cost air flow control, and for residential applications of LMB, hot air could be re-used.

The paper is organized as follows: in the next section, we formulate the equations governing the heat transfer in LMBs and give the boundary conditions, with nonlinear dependency on the absolute temperature. In the section III, the temperature profiles in different liquid layers are determined and operating regimes of the LMB are determined as function of the aspect ratio. Thermo-convective flows in the LMB are determined using finite elements from COMSOL Multiphysics. The last section contains discussions and concluding remarks.

## II. Problem formulation

We consider a cylindrical battery made of two liquid layers of electrodes that sandwich an electrolyte layer (Fig.1). The electrodes are made of liquid metals A and B, and the electrolyte E is often a molten salt. The battery has length $H$ and radius $a$ yielding an aspect ratio $\Gamma = \frac{H}{a}$. The liquids have the following properties: density $\rho_i$, thermal expansion coefficient $\alpha_i$, kinematic viscosity $\nu_i$, thermal diffusivity $\kappa_i$, specific heat capacity $c_{pi}$ and electrical conductivity $\sigma_i$, where the index $i$ refers to the liquid layers A, E and B. The density of the liquid is chosen to ensure stable density stratification in the gravity field i.e. $\rho_A < \rho_E < \rho_B$.

An imposed uniform vertical electric current $I$ of constant density $\vec{j} = j_0 \vec{e}_z$ i.e. $j_0 = I/(\pi a^2)$ crosses vertically the liquid layers, resulting in internal Joule heating with a generated power density $\dot{q}_i = j_0^2/\sigma_i$. As we are interested in the conduction state of the liquid layers in the LMB, the Lorentz force induced by $j_0$ will be ignored as it stabilizes the convective motion in the conducting liquid [19].

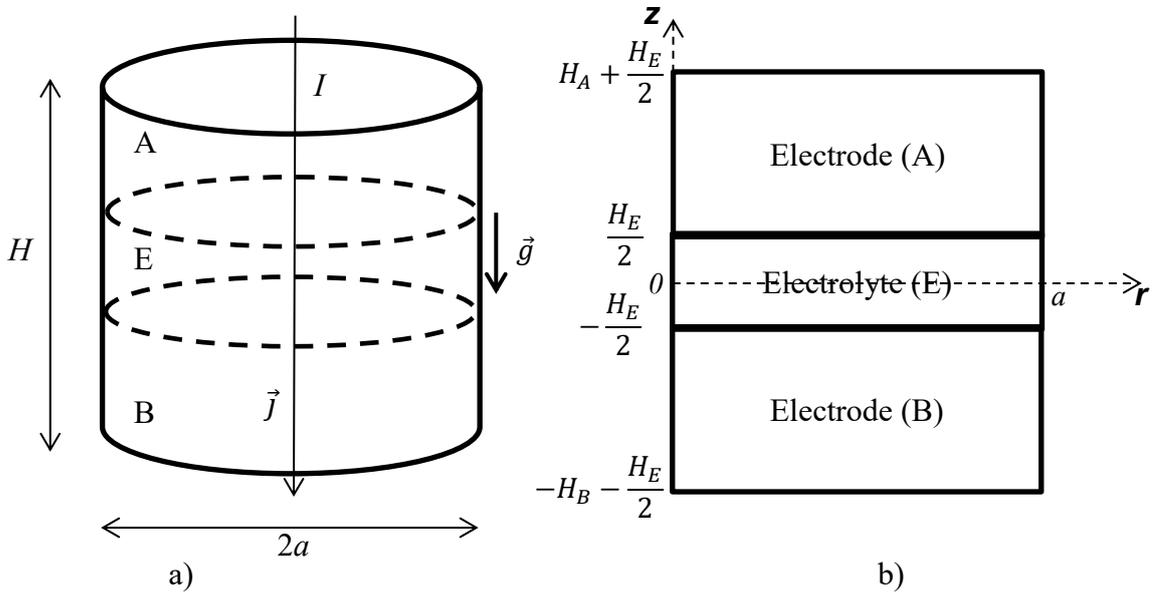

Figure 1: a) cylindrical Liquid metal battery: b) meridional half cross-section.

*a. Governing equations*

The internal Joule heating in the battery generates temperature gradients in the liquid layers and therefore convective flows can be induced above critical values of the temperature gradients [13-16]. In order to characterize the temperature distribution and the induced flows in the battery, we write the Navier-Stokes equations coupled with the energy equation in each liquid layer of the battery [5, 7]:

$$\vec{\nabla} \cdot \vec{u}_i = 0 \qquad (1\text{-a})$$

$$\frac{\partial \vec{u}_i}{\partial t} + (\vec{u}_i \cdot \vec{\nabla})\vec{u}_i = -\frac{1}{\rho_{oi}}\vec{\nabla} p_i + \nu_i \vec{\nabla}^2 \vec{u}_i - \alpha_i \theta_i \vec{g} \qquad (1\text{-b})$$

$$\frac{\partial \theta_i}{\partial t} + (\vec{u}_i \cdot \vec{\nabla})\theta_i = \kappa_i \vec{\nabla}^2 \theta_i + \frac{j_0^2}{(\rho_0 c_p \sigma)_i} \tag{1-c}$$

where $i \in \{A, E, B\}$, $\vec{u}_i$, $p_i$ and $\theta_i = T_i - T_0$ are the velocity, the pressure and the deviation of the temperature from the ambient temperature $T_0$, $\rho_i = \rho_{0i}(1 - \alpha_i \theta_i)$ is the density, $\vec{g}$ is the gravity acceleration. We have assumed the Oberbeck-Boussinesq approximation, i.e. all the physical properties are constant except in the buoyancy term. For electrolytes used in the LMB, $\sigma_E \ll (\sigma_A, \sigma_B)$, hence the rate of volumetric Joule heating in the liquid metals (A and B) can be neglected compared to that in the electrolyte. The Joule heating will generate a convective flow in the electrolyte, and the top electrode may become unstable because of the negative temperature gradient.

*b. Boundary conditions*

There are two types of boundary conditions: boundary conditions at the solid wall of the battery, and the matching conditions at the interfaces A/E and E/B. The no-slip condition at the walls of the container:

$$u_i(r = a) = u_A(z = \tfrac{1}{2}H) = u_B(z = -\tfrac{1}{2}H) = 0 \tag{2-a}$$

The kinematic conditions at the interfaces concern the continuity of the velocity components i.e.

$$u_{r,A} = u_{r,E};\ u_{\theta,A} = u_{\theta,E};\ u_{z,A} = u_{z,E}\ \text{at}\ z = \tfrac{1}{2}H_E \tag{2-b}$$

$$u_{r,B} = u_{r,E};\ u_{\theta,B} = u_{\theta,E};\ u_{z,B} = u_{z,E}\ \text{at}\ z = -\tfrac{1}{2}H_E \tag{2-c}$$

The dynamic conditions are given only by the continuity of stresses at the interfaces, these being assumed to remain flat because of the large values of the interface tension between liquid metals and electrolytes [7, 20]:

$$2\eta_E \frac{\partial u_{z,E}}{\partial z} - 2\eta_A \frac{\partial u_{z,A}}{\partial z} = p_E - p_A\ \text{at}\ z = \frac{H_E}{2} \tag{3-a}$$

$$\eta_A \left(\frac{\partial u_{r,A}}{\partial z} + \frac{\partial u_{z,A}}{\partial r}\right) - \eta_E \left(\frac{\partial u_{r,E}}{\partial z} + \frac{\partial u_{z,E}}{\partial r}\right) = 0\ \text{at}\ z = \frac{H_E}{2} \tag{3-b}$$

$$\eta_A \left(\frac{\partial u_{\theta,A}}{\partial z} + \frac{1}{r}\frac{\partial u_{z,A}}{\partial \theta}\right) - \eta_E \left(\frac{\partial u_{\theta,E}}{\partial z} + \frac{1}{r}\frac{\partial u_{z,E}}{\partial \theta}\right) = 0\ \text{at}\ z = \frac{H_E}{2} \tag{3-c}$$

$$2\eta_B \frac{\partial u_{z,B}}{\partial z} - 2\eta_E \frac{\partial u_{z,E}}{\partial z} = p_B - p_E\ \text{at}\ z = -\frac{H_E}{2} \tag{3-d}$$

$$\eta_E \left(\frac{\partial u_{r,E}}{\partial z} + \frac{\partial u_{z,E}}{\partial r}\right) - \eta_B \left(\frac{\partial u_{r,B}}{\partial z} + \frac{\partial u_{z,B}}{\partial r}\right) = 0\ \text{at}\ z = -\frac{H_E}{2} \tag{3-e}$$

$$\eta_E \left( \frac{\partial u_{\theta,E}}{\partial z} + \frac{1}{r} \frac{\partial u_{z,E}}{\partial \theta} \right) - \eta_B \left( \frac{\partial u_{\theta,B}}{\partial z} + \frac{1}{r} \frac{\partial u_{z,B}}{\partial \theta} \right) = 0 \text{ at } z = -\frac{H_E}{2} \qquad (3\text{-f})$$

where $\eta_i = (\rho \nu)_i$ is the dynamic viscosity of the layer $i$.

The thermal boundary conditions at the interfaces are the continuity of the temperature and heat fluxes:

$$T_A = T_E \quad \text{and} \quad \lambda_A \frac{\partial T_A}{\partial z} = \lambda_E \frac{\partial T_E}{\partial z} \text{ at } z = \frac{1}{2} H_E \qquad (4\text{-a})$$

$$T_E = T_B \quad \text{and} \quad \lambda_E \frac{\partial T_E}{\partial z} = \lambda_B \frac{\partial T_B}{\partial z} \text{ at } z = -\frac{1}{2} H_E \qquad (4\text{-b})$$

where $\lambda_i = (\rho c_p \kappa)_i$ is the thermal conductivity of the liquid layer $i$. The thermal boundary conditions at the electrodes with the ambient medium are the Fourier-Robin conditions [20]:

$$-\lambda_A \frac{\partial T_A}{\partial z} = h_A (T_A - T_0) \text{ at } z = \frac{1}{2} H \qquad (4\text{-c})$$

$$\lambda_B \frac{\partial T_B}{\partial z} = h_B (T_B - T_0) \text{ at } z = -\frac{1}{2} H \qquad (4\text{-d})$$

where $h_A$ and $h_B$ are the overall heat exchange coefficients associated with both the electrodes. They are defined through the total outward flux of heat transfer which is the sum of the convective and radiative heat transfers, i.e. $h(T_{sk} - T_0)$ given by Holman [22]:

$$h_k = h_g + \varepsilon \sigma (T_{sk} + T_0)(T_{sk}^2 + T_0^2) \qquad (5\text{-a})$$

where $k \in \{A, B\}$, $h_g$ is the convective heat transfer, $\varepsilon$ is the emissivity of the outer surface of the battery container, $\sigma = 5.6704 \times 10^{-8}$ W/(m²K⁴) is the Stefan-Boltzmann constant, $T_{sk}$ ($k \in \{A, B\}$) is the temperature of the outer surface of the electrode $k$:

$$T_{sA} = T_A (z = \frac{1}{2} H) \qquad (5\text{-b})$$

$$T_{sB} = T_B (z = -\frac{1}{2} H) \qquad (5\text{-c})$$

Finally at the radial boundary the LMB container is insulated:

$$-\lambda_i \frac{\partial T_i}{\partial r} = 0 \text{ at } r = a. \qquad (6)$$

# III. Results

**a. Base state of the LMB**

In this state, there is no flow in the LMB and the Joule heating induces a conducting state with only stationary vertical temperature dependence, i.e. $T_i = T_i(z)$, which satisfies the conduction equation:

$$\frac{d^2 T_i}{dz^2} = -\frac{j_0^2}{\lambda_i \sigma_i} \qquad (7)$$

The electrolyte has small electric and thermal conductivities compared to the electrodes so that the source term in the equation (7) can be neglected in the electrodes. Then, the temperature profile has a quadratic shape in the electrolyte while it is linear in the electrode layers. The solution to equations (4) and (7) reads

$$T_A = T_0 + \frac{H_E j_0^2}{\sigma_E}\left(h_A^{-1} + R_A \frac{-z + H_A + \frac{1}{2}H_E}{H_A}\right)\left[\left(\frac{R_E}{2} + R_B + h_B^{-1}\right)/R_{tot}\right] \qquad (8\text{-a})$$

$$T_E = T_0 + \frac{H_E j_0^2}{\sigma_E}\Bigg\{-\frac{1}{2}R_E\left(\frac{z}{H_E}\right)^2$$
$$+ \frac{R_E}{2R_{tot}}\left[\frac{z}{H_E}(R_A + h_A^{-1} - R_B - h_B^{-1}) + \frac{3}{4}\left(R_{tot} - \frac{2}{3}R_E\right)\right. \qquad (8\text{-b})$$
$$\left.+ \frac{2}{R_E}(R_A + h_A^{-1})(R_B + h_B^{-1})\right]\Bigg\}$$

$$T_B = T_0 + \frac{H_E j_0^2}{\sigma_E}\left(h_B^{-1} + R_B \frac{z + H_B + \frac{1}{2}H_E}{H_B}\right)\left[\left(\frac{R_E}{2} + R_A + h_A^{-1}\right)/R_{tot}\right] \qquad (8\text{-c})$$

where we have introduced the thermal resistances $R_i = H_i/\lambda_i$ together with the total thermal resistance $R_{tot} = h_A^{-1} + R_A + R_E + R_B + h_B^{-1}$. Although the highest temperature is not reached at $z = 0$, the total available heat flux $H_E \dot{q}_E = H_E j_0^2/\sigma_E$ is split between upward and downward fluxes, according to the relative thermal resistances at $z \leq 0$ and $z \geq 0$ respectively. The equations (8-a,b,c) contain the dependence on surface temperature $T_{sk}$ included in the heat transfer coefficients $h_A$ and $h_B$ in equations (5). This results in asymmetric values $h_A \neq h_B$ if $R_A \neq R_B$.

One can distinguish two cases: i) symmetrical battery with $\lambda_A = \lambda_B$, $h_A = h_B$, $H_A = H_B$ which represents a simplified hypothetical model investigated by Shen and Zikanov [5] with $h_A = h_B = \infty$ and, ii) the asymmetrical battery for which at least one of the following conditions is satisfied $\lambda_A \neq \lambda_B$, $h_A \neq h_B$, $H_A \neq H_B$. When $(h_A, h_B) \rightarrow \infty$, that is isothermal outer surfaces, we retrieve the profiles used by Shen and Zikanov [5] for symmetrical LMB, and by Bradwell et al. [2] or Köllner et al. [7] for asymmetrical battery ($\lambda_A \neq \lambda_B$). From the expressions of the temperature distributions (8-a,b,c), we determine the minimum and the maximum temperature in each liquid layer of the LMB:

$$T_{A,min} = T_{sA} = T_A(z = \tfrac{H_E}{2}) \; ; \; T_{A,max} = T_A(z = \tfrac{H_E}{2}) \tag{9-a}$$

$$T_{E,max} = T_E(z = \tfrac{1}{2}H_E[R_A + h_A^{-1} - R_B - h_B^{-1}]/R_{tot}) \tag{9-b}$$

$$T_{B,min} = T_{sB} = T_B(z = -\tfrac{H_E}{2}) \; ; \; T_{B,max} = T_B(z = -\tfrac{H_E}{2}) \tag{9-c}$$

The temperature of the conduction state for different boundary conditions can be expressed in terms of the reduced temperature:

$$\theta^*(z^*) = [T(z) - T_{sB}] / \Delta T \tag{10}$$

where $\Delta T = \tfrac{1}{4}R_E H_E \dot{q}_E = (\tfrac{1}{2}H_E j_0)^2/(\sigma_E \lambda_E)$, and $z^* = 2z / H_E$

The temperature profiles $\theta^*(z^*)$ of the asymmetrical LMB with and without the Fourier-Robin boundary conditions are illustrated in Figure 2, for a Ca-CaCl$_2$LiCl-Sb LMB. For such battery, $R_A$ is much lower than the other thermal resistances, resulting in a flatten temperature profile in the top electrode, and a temperature difference between $T_{sA}$ and $T_{sB}$.

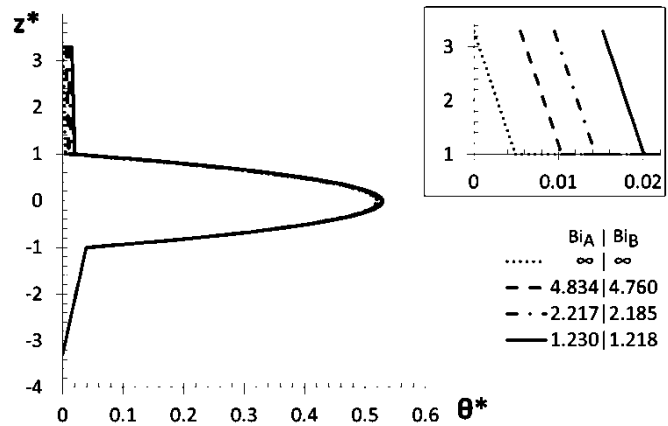

Figure 2: Vertical profile of the reduced temperature in the conducting state for various Biot numbers $Bi_k = h_k R_k$, $\in \{A, B\}$ ; for a Ca-CaCl$_2$LiCl-Sb LMB where $\Gamma = 1$, $R_A/R_E = 2.41 \times 10^{-3}$, $R_B/R_E = 19.81 \times 10^{-3}$, $\Delta T = 932.7$ K. Inset: zoom of reduced temperature profiles in the upper layer.

However when the heat exchange coefficients decrease, the imbalance between $(R_A + h_A^{-1})/R_{tot}$ and $(R_B + h_B^{-1})/R_{tot}$ also decreases, as $h_A^{-1}$ increases and dominates $R_A$. The temperature difference between $T_{sA}$ and $T_{sB}$ is then reduced.

**b. Convective motion in the LMB**

We have solved numerically the governing equations 1, in all layers, using the interfacial fluid-fluid feature of the finite elements software COMSOL MULTIPHYSICS. A large surface tension $\sigma = 100$ N/m was used to obtain negligible deformation of interfaces. The duration of the simulations was chosen to be $100\ \tau_{conv}$ where the convection timescale $\tau_{conv}$ is defined as $\tau_{conv} = \sqrt{a/(\alpha \Delta T g)}$. Our numerical simulations were performed for a LMB with symmetric properties [5], $\Gamma = 1$, $\gamma = H_A/H = H_B/H = 0.348$, $\lambda_A = \lambda_B$, and imposed symmetric heat transfer coefficients $h = h_A = h_B$ such that the Biot numbers Bi $= hR_E = 1.352$. For a battery with a volume $V = 0.0205$ m³ and applied current $j_0 = 10^4$ A/m², the parameters in [5] result in $\Delta T = 932.7$ K and $\tau_{conv} = 0.264$ s.

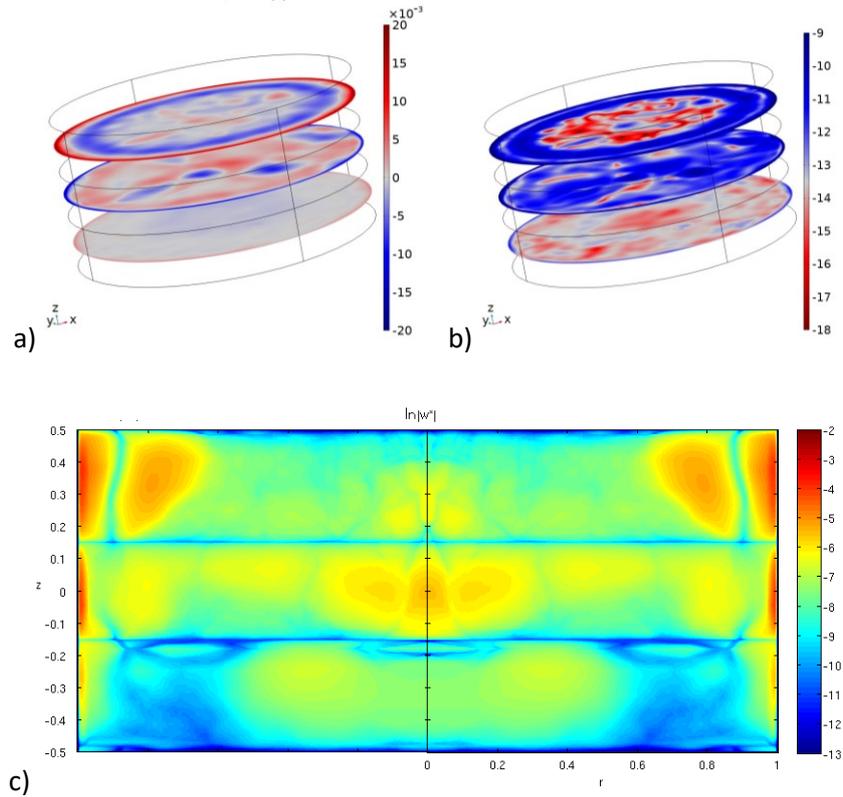

Figure 3: Visualization of the field in all three layers at $t = 100 \cdot \tau_{conv}$, Bi $= 1.352$, $(a^2/\nu)/\tau_{conv} = 1.86 \cdot 10^5$ : a) the vertical velocity component, b) kinetic energy (*logarithmic* scale), c) vertical component of the velocity averaged in the azimuthal direction (displayed in *logarithmic* scale ; the left part of the image is added by symmetry as a visual guide). $\Gamma = 1$, $\gamma = 0.348$.

The cross-sections of the vertical velocity and of the kinetic energy at the center of each liquid layer are illustrated in Figure 3. Both the quantities take significant values in the electrolyte and in the upper electrode but are negligible in the lower electrode, where we observe a weak convective flow (Fig. 3c) due to the anticonvection flow [9, 22] compared to the convective flow in the top electrode. Indeed, the ratio of maximal velocities in both the electrodes are $|w^*|_{max,A}/|w^*|_{max,E} = 1.175$ and $|w^*|_{max,B}/|w^*|_{max,E} = 0.250$.

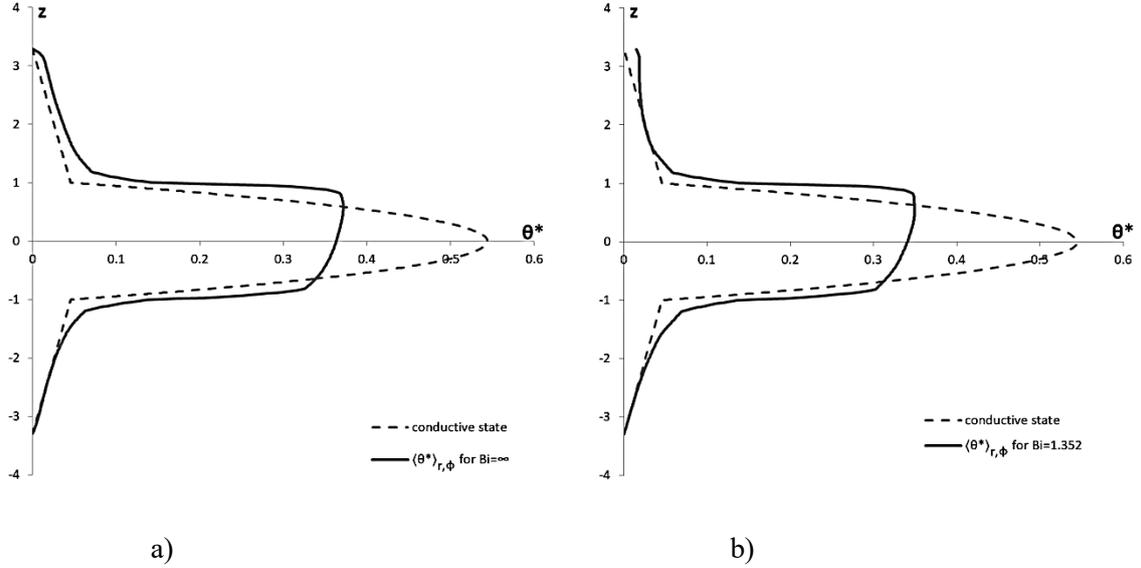

a) b)

Figure 4: At $t = 100 \cdot \tau_{conv}$, mean temperature $\langle \theta^* \rangle_{r,\phi}$ as a function of $z^*$ in the conductive (dotted line) and convective regimes (continuous) with: (a) Bi $= \infty$, and (b) finite heat exchange coefficients Bi $= 1.352$.

The maximum value of temperature in convection state is lower than that of the conduction in the electrolyte (Fig. 4), as already reported [7]. Figure 4 illustrates that in the convective regime for both isothermal and Fourier-Robin conditions, there is almost no change in the temperature profiles in the electrodes except near the interfaces with the electrolyte where thermal boundary layers develop. For finite value of $h$, the temperature profile in the top layer is flattened, and, as a symmetric LMB is considered here, no imbalance exists between $(R_A + h_A^{-1})$ and $(R_B + h_B^{-1})$. Thus the close-to-flat profile is the result of enhanced heat transfer by convection. However, the reached temperature range still compares well with the conductive state. Hence the prediction of thermally-operating regime can be based on the parameters of the conduction state.

### c. Thermally-Operating Regime of the LMB

The electrolyte being a molten salt (liquid state), we define the thermally-operating regime of the LMB by the constraint that the electrodes, after they fuse, should remain in the liquid state when the current crosses them, i.e. they should remain below the boiling point $T_{k,b}$, $k \in \{A, B\}$. Thus, we impose the following conditions to the working temperatures of the battery:

$$T_{k,melt} < T_{k,min} \; ; T_{k,max} < T_{k,boil}; \quad k \in \{A, B\} \tag{11}$$

The maximum and minimum temperatures $T_{k,min}$; $T_{k,max}$ of the electrode layers in the LMB depend on the applied density current $j_0$ and the layers thicknesses $H_i$, $i \in \{A, E, B\}$, in the LMB. We assume that dissociation does not take place in the electrolyte.

c.1. Estimate of the charge/discharge time of the electrode A

The choice of the applied density current $j_0$ is determined by the characteristic time of charge/discharge of the LMB with a given aspect ratio $\Gamma = H/a$ and a fixed volume $V = \pi a^2 H$. The characteristic charge/discharge time of the electrode A is connected to its volumetric capacity $C_A$ by the Peukert relation [23] $C_A \cdot V_A = t(Sj)^k$ which yields for $k = 1$ the longest characteristic time of charge/discharge of the LMB: $t_{cd} = H\gamma C_A/j$ where $\gamma = H_A/H$ is the ratio of the thickness of the upper electrode to the total height of the LMB, $C_A = zeN_{Av} \rho_A/M_W$ with $N_{Av}$ the Avogadro number, $z$ and $M_W$ being the valence and atomic weight of the electrode element respectively. As a first estimate for $t_{cd}$ we use the value of $\gamma$ and $\Gamma$ from the LMB described by [2] but for $\gamma = 0.348$, $\Gamma = 0.5$, and get $t_{cd} = 7.7$ h for $j_0 = 10^4$ A/m² and a volume $V = 0.0205$ m³.

c.2. Dependence on the aspect ratio

The relations (9) show that the dependence of $T_{i\,max,min}$ on the height of the container $H$ can be expressed in terms of the aspect ratio $\Gamma$ for a fixed total volume $V = \pi a^2 H$ of the battery. Indeed, the height $H$ is related to the aspect ratio $\Gamma$ and the volume $V$ of the battery through the relation $H = (V/\pi)^{1/3} \Gamma^{2/3}$. We have plotted in Figure 5 the dependence of temperature $T_{i\,max,min}$, $i \in \{A, E, B\}$ obtained from (9) on the aspect ratio $\Gamma$. We also have added the melting and boiling temperature of both the electrodes. When the heat exchange is infinite, there is no operating regime of the LMB for any reasonable value of the aspect ratio (Fig.5-a). A thermally-operating regime is found for finite value of exchange coefficient in the LMB (Fig.5-b) with aspect ratio $\Gamma \in [0.22, 0.82]$. These values were obtained for a current density of $j_0 = 10^4 \text{A/m}^2$. The estimated characteristic time of charge/discharge would then range in the interval [4.5, 10.7] hours. If the volume is increased, the admissible aspect ratio is decreased.

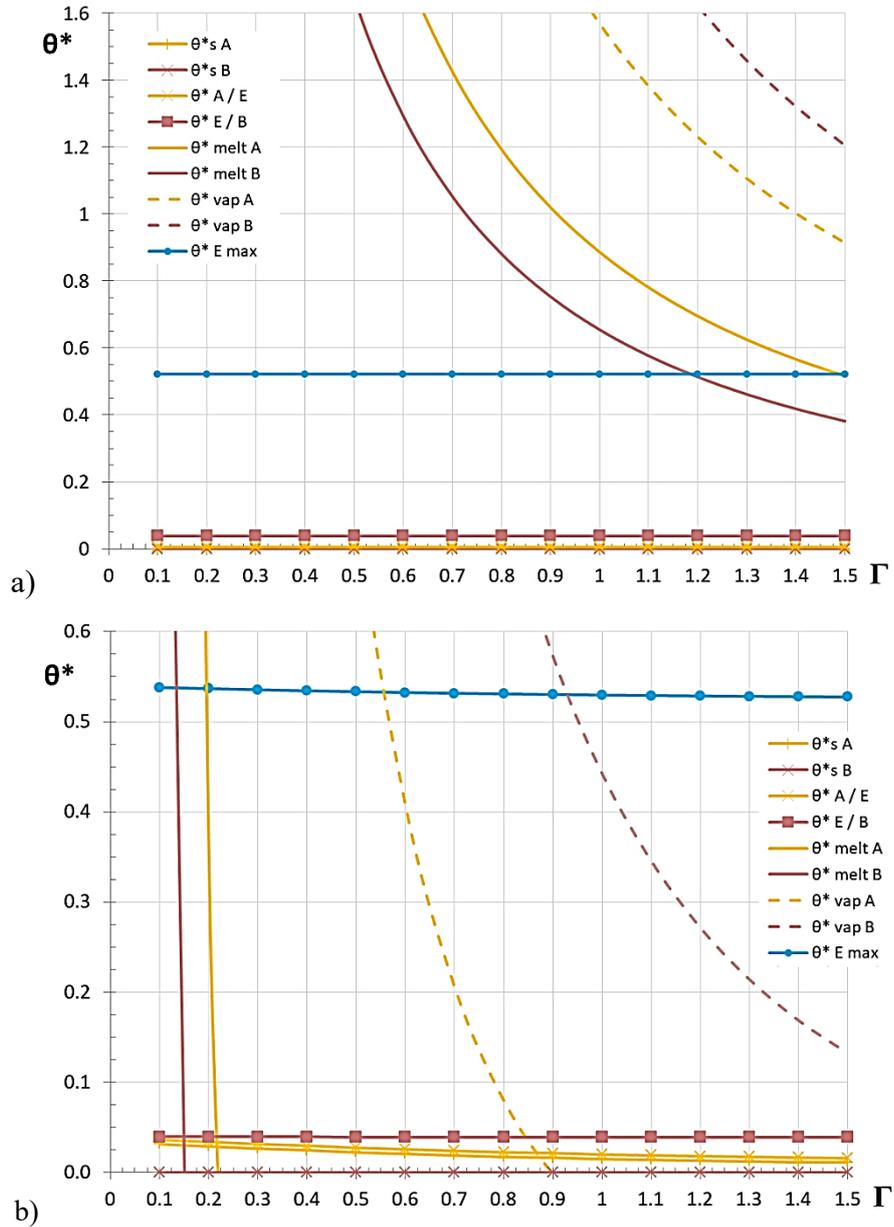

Figure 5: Variation of the minimum $T_{min}$ and maximum $T_{max}$ displayed on the diagram $(\Gamma, \theta^*)$ for the battery Ca-CaCl$_2$LiCl-Sb [1-3], $\gamma = 0.348$, and a fixed volume $V = 0.0205$ m³. a) Bi $= \infty$, b) Bi$_A$ = 1.230, Bi$_B$ = 1.218.

## IV. Discussion

In the present study, we have investigated the heat transfer induced by the internal heating in the electrolyte in the whole battery. A specific focus was made on the effect of heat exchange between the battery and its environment. We have highlighted the effect of the temperature increase due to the finite value of the heat coefficient; we retrieved the results obtained by Shen and Zikanov [5] and of Köllner *et al.* [7] when the heat exchange with the environment is infinite i.e. when the boundary of the container are supposed to be at the same temperature as the environment.

The convective regime has been simulated and it shows that the temperature profile taking into account the heat exchange coefficient is flatter in the convective flow than the one obtained with infinite heat exchange coefficient [5, 7]. This demonstrates a local enhancement of the heat transfer, in particular in the upper electrode, although thermal equilibrium is still obtained (all the heat generated in the electrolyte leaves the battery, as verified in our numerical simulation). However the highest temperatures reached in the battery as a whole, in the electrolyte, are comparable when considering the reduced temperature. Hence, when the heat exchange coefficients are taken into account, the effect on heat transfer performance is mostly summarized by the temperature jumps at the boundaries, and the reduced imbalance between the upward and the downward thermal paths.

The maximum temperature in the convection regime in the electrolyte is lower than in the conduction, while in the electrodes the temperature does not differ significantly from that of the conduction regime. So the estimate of the thermally-operating regime can be based on the temperatures of the conduction state and the working materials remain in the liquid state. For a fixed volume of the battery, an operating regime of the allowed temperatures needed to maintain the electrodes and electrolyte in the liquid states has been established depending of the aspect ratio of the battery.

Phenomena such as the electro-vortex flow, the Tayler instability, the mass transfer or the deformability of the interface have been addressed in recent studies by different authors [7-12, 24-26]. The electric current across the liquid in a vertical direction $\vec{j} = j_0 \vec{e}_z$ generates an azimuthal magnetic field $\vec{B} = B(r)\vec{e}_\phi$ where $B(r)$ is given by:

$$B(r) = \begin{cases} \mu j_0 r/2, & r < a \\ \mu j_0 a^2/2r, & r > a \end{cases}$$

with $\mu_0 = 4\pi \times 10^{-7}$ H/m is the magnetic permeability of the vacuum. The magnetic field reaches its maximum value at the surface of the container, i.e. $B_{max} = \mu j_0 a/2$. For the available models of the LMB [1-3], the current density is of the order $j_0 \sim 10^3 - 10^4$ A/m² and the radius $R \sim 0.1$ m so that $B_{max} = 6.2832 \times 10^{-5} - 10^{-4}$ T. This field can be compared to the characteristic magnetic of the liquid metals $B_i = (\rho v/\sigma H_i^2)^{1/2}$. In the commonly used electrodes in LMB, $B_i \sim 10^{-4}$ T and much higher in the electrolyte. The Hartmann number $\text{Ha} = B_{max}/B_i \sim 10^{-1} - 10^0$ associated with the induced magnetic field by the current crossing the LMB is too small to modify the convection [4, 19, 27, 28]. One should mention that internal heating-induced convection in the LMB can be delayed by an applied magnetic field which is known to dissipate flow energy [19, 27, 28]. This suggests that properly chosen imposed magnetic field can maintain the liquid layers in the LMB in a conduction state. A thorough study of combined effect of an internal-heating convection and an applied magnetic field is needed to improve the thermally-working regime of the LMB.

Recent studies have made focus on different phenomena susceptible to develop in a LMB such as the electrothermal heating [6] which can modify the present results, the deformability of the interface [26], the Tayler instability [12] or the electro-vortex flow [24]. In particular, we can estimate the ratio of the characteristic velocities of electro-vortex flow to internally heating-induced convection [24] $A_i = \frac{a}{2H_i}\left(\frac{\mu_0 \sigma \lambda}{\rho g \alpha H_i}\right)^{1/2}$. In the electrolyte LiCl-KCl of the LMB investigated by Köllner *et al.* [7], the parameter $A_i' = \frac{2H_i}{a}A_i \sim 10^{-2}$ for $H_i = 10^{-2}$ m suggesting that the electro-vortex flow is dominated by the internal heating-induced convection in the electrolyte.

## V. Conclusion

The present study concerns the determination of the thermally-operating of the liquid metal battery. The heat exchange coefficient between the battery and the environment has been taken into account while the magnetic field effects have been neglected. The heat exchange coefficients lead to more symmetric split of the heat flux between top and bottom electrodes. The applied current density and geometry can be chosen such that the characteristic time of charge/discharge compares with half a day. This prediction can be based on purely conductive temperatures in the electrodes as, although convection can increase heat transfer in the top electrode, it does not influence sufficiently the highest and lowest temperatures in this electrode.

The effects of convection in shallower electrolyte layers remain to be investigated, and may require lower exchange coefficient than presently investigated to maintain thermally operating conditions. Also LMBs with a void at the top may be better described by taking into account an heat exchange coefficient.

**Acknowledgments**

The present work has benefited from the financial support from the French National Research Agency (ANR) through the program Investissements d'Avenir (ANR-10 LABX-09-01), Labex EMC$^3$ (project HILIMBA). R.H. was partially supported by the graduate school MES (Materials and Energy Sciences).